\documentclass[nofootinbib,twocolumn,pre,superscriptaddress]{revtex4-1}

\makeatletter
\newif\if@restonecol
\makeatother

\usepackage[ruled,vlined]{algorithm2e}
\usepackage[usenames]{xcolor}
\usepackage{balance}
\usepackage{graphicx}
\usepackage{amsmath}

\begin{document}

\title{Predicting sports scoring dynamics with restoration and anti-persistence}

\author{Leto Peel}
\email{leto.peel@colorado.edu}
\affiliation{Department of Computer Science, University of Colorado, Boulder, CO 80309}
\author{Aaron Clauset}
\email{aaron.clauset@colorado.edu}
\affiliation{Department of Computer Science, University of Colorado, Boulder, CO 80309}
\affiliation{BioFrontiers Institute, University of Colorado, Boulder, CO 80303}
\affiliation{Santa Fe Institute, 1399 Hyde Park Rd., Santa Fe, NM 87501}

\begin{abstract}
Professional team sports provide an excellent domain for studying the dynamics of social competitions.  These games are constructed with simple, well-defined rules and payoffs that admit a high-dimensional set of possible actions and nontrivial scoring dynamics.  The resulting gameplay and efforts to predict its evolution are the object of great interest to both sports professionals and enthusiasts. 
In this paper, we consider two online prediction problems for team sports:~given a partially observed game~\textit{Who will score next?} and ultimately~\textit{Who will win?} We present novel interpretable generative models of within-game scoring that allow for dependence on lead size (\textit{restoration}) and on the last team to score (\textit{anti-persistence}). We then apply these models to comprehensive within-game scoring data for four sports leagues over a ten year period. By assessing these models' relative goodness-of-fit we shed new light on the underlying mechanisms driving the observed scoring dynamics of each sport.  Furthermore, in both predictive tasks, the performance of our models consistently outperforms baselines models, and our models make quantitative assessments of the latent team skill, over time.  

\end{abstract}

\maketitle

\section{Introduction}
Competition in social systems is a natural and pervasive mechanism for improving performance and distributing limited resources.  The quantitative study of such competitions can improve our ability to predict the outcomes associated with specific strategies and the strategic choices that competitors may make. However, most real competitions take place in complex and evolving environments~\cite{galla2013complex,merritt2013environmental}, which makes them difficult to study. Professional team sports, with their well defined and consistently enforced rules, provide a controlled setting for the study of competition dynamics~\cite{reed2006an,gabel2012random,merritt2014scoring} and have previously been used as model systems for studying business decision making and human behavioral biases~\cite{rabin2010gambler, stockl2013hot}. 
The recent trend toward recording comprehensive and detailed data on the events within particular games provides us with new opportunities to study, model, and predict the dynamics of these games.  The results of these studies promise to shed new light on a wide variety of existing competitive social systems, and enhance work on designing new ones, both offline and online.

Here, we examine the time series of scoring events in all league games across four different team sports over a period of ten years.  We construct and test probabilistic models for two online predictive tasks: given a partially observed game~\textit{Who will score next?}~and ultimately~\textit{Who will win?} We then use these models to investigate the predictiveness of the dynamical phenomena of \textit{restoration} and \textit{anti-persistence}, which are defined below. 

The events within a particular game can be effectively modeled as the interaction of skill and chance.  Inferring skill from a series of competitions has a long history of study, both for individuals~\cite{elo1978rating, glickman1999parameter} and for teams~\cite{Herbrich2007trueskill,tarlow2014knowing}. However, this past work has typically only considered the final outcome of games, in terms of either a win or loss, or the final point difference. Here, we focus on modeling the specific pattern of scoring events within an individual game.

The role of chance also has a long history of study, typically focusing on the question of whether one success increases the likelihood of subsequent success.  This idea can be formalized at different levels, e.g., success by individual players within a game~\cite{bar2006twenty, gilovich1985hot,yaari2011hot}, or a team's success across multiple games~\cite{arkes2011finally,sire2009understanding,vergin2000winning}. Here, for the first time, we focus on a different level: success by a whole team within a game.

A simple starting point for such models is the basic idea of many skill ranking systems~\cite{bradley1952rank, elo1978rating}, which model game outcomes as random variables dependent on the competing teams' skills. We extend this idea to consider the point-scoring events within a game to be a sequence of independent contests. Past work supports this approach, as some studies have found a lack of dependence between an individual scoring and their ability to score subsequent points~\cite{gilovich1985hot,yaari2011hot}, or between a team winning and their chance to win future games~\cite{sire2009understanding,vergin2000winning}.  
On the other hand, there is also evidence of non-independence, e.g., the probability of scoring itself can vary with the clock time within a game or with the size of the lead~\cite{gabel2012random, merritt2013environmental, merritt2014scoring}. To investigate the degree to which non-independence governs scoring probabilities, we construct a sequence of more complex models that allow specific aspects of a game's current state to influence scoring rates, e.g., the team that scored last and the lead size.

In many sports, including American football and basketball, a simple source of non-independence is a forced change in ball possession after each scoring event, putting the scoring team at a disadvantage.  This can result in a phenomenon called \textit{anti-persistence}, in which a score by one team is more likely to be followed by a score by their opponent~\cite{gabel2012random}.

Another potential source of non-independence is the size of the lead itself. Past work has shown that the observed probability of scoring next can vary with lead size~\cite{gabel2012random, merritt2014scoring}. 
A negative dependence may be the result of strategy, e.g., a team using its best players when it falls behind and substituting them out when they are ahead. Such strategies have a \textit{restorative} effect on the lead size, serving to pull the size of the lead back toward zero. Conversely, anti-restoration or momentum occurs when the leading team has a higher chance of scoring again, perhaps by improving their control over the playing field or by learning from gameplay how to better exploit the weaknesses of the opposing team.

In this paper, we develop probabilistic generative models around these ideas to explore and predict the evolution of point scoring over the course of a game. We use these models to deduce the impact of chance, strategy, and the rules of the game itself, and to test two simple hypotheses:
\begin{enumerate}
\itemsep-1.0pt 
  \item the probability of scoring \textit{does not} depend on the current state of the game (team skill alone matters).
  \item the probability of scoring \textit{does} depend on the current game state (as well as team skill).
\end{enumerate}

Our probabilistic models encode specific instances of these assumptions and we assess their accuracy under two online predictive tasks. We present novel predictive models that can not only predict the outcome of a game, but also provide better predictions over baseline models about the sequence of scoring events.  

\section{Related Work}
Our work addresses two novel prediction problems for predicting \textit{Who will score next?}~and \textit{Who will win?}, using only the sequence of scoring events that have already occured during the game.  In the following we outline related work to each of these questions in turn. 

Essential to answering the question \textit{Who will score next?}~is understanding the underlying mechanisms of scoring dynamics.  The study of competitive team sports has a rich history spanning a broad selection of features including the timing of scoring events~\cite{buttrey2011estimating, everson2008composite,gabel2012random,heuer2010soccer, merritt2014scoring,thomas2007inter, yaari2011hot}, long-range correlations in scoring~\cite{ribeiro2012anomalous}, the role of timeouts~\cite{saavedra2012is}, the identification of safe leads~\cite{clauset2015safe}, and the impact of spatial positioning and playing field design~\cite{bourbousson2012space, merritt2013environmental, yue2014learning}.  
The most relevant of these studies focuses on the analysis of individual player ``momentum'' or ``hot-hands''~\cite{bar2006twenty, gilovich1985hot,yaari2011hot} and on team winning streaks~\cite{arkes2011finally, gilovich1985hot, sire2009understanding, vergin2000winning, yaari2011hot}. Here, we bring together these two ideas by considering the notion of momentum, or its reverse ``restoration'', at the team level.  Although some analysis has previously been undertaken in this direction~\cite{gabel2012random}, we go further to provide the first predictive models that answer the question:\ \textit{Who will score next?}

The foundations of our approach lie in the field of skill modeling and team ranking~\cite{bradley1952rank,elo1978rating}, which originated in the mid-20th century. 
Work in this area includes the ranking of individuals~\cite{dangauthier2007trueskill, glickman1999parameter, elo1978rating}, teams~\cite{glickman1998state, sire2009understanding, tarlow2014knowing}, or both~\cite{Herbrich2007trueskill, huang2006generalized}.  These models have been applied to a wide range of competitive events, including baseball~\cite{sire2009understanding}, chess~\cite{dangauthier2007trueskill, glickman1999parameter, elo1978rating}, American football~\cite{glickman1998state, tarlow2014knowing}, association football (soccer)~\cite{hvattum2010using}, and tennis~\cite{glickman1999parameter}.  More recently, they have been adapted to matchmaking problems in online games~\cite{Herbrich2007trueskill} and to calibrating reviewer scores in computer science conferences~\cite{flach2010novel}.

Our work is the first to use skill ranking models to predict \textit{Who will win?}~by predicting the sequence of scoring events within a game.  Skill ranking models have previously been applied to predicting game outcomes but only based on the final outcome of the game, either in terms of the win/loss result or the final point difference.  These past approaches thus cannot update their prediction as the game unfolds, while our models can.  We train on a history of scoring event sequences so that we may predict \textit{Who will win?}~in an online fashion.  Some commercial online sports betting systems exist that make similar online predictions, but these systems are proprietary and closed, which precludes a scientific evaluation or comparison with our models. They are not considered hereafter.

\begin{table*}
\centering
  \caption{Summary of our sports data for multiple seasons across four team competitive sports.}
\label{table:game_stats}
\begin{tabular}{l|l|c|r|rr|rr|r}
 &  & & & \multicolumn{2}{c|}{number of games} & \multicolumn{2}{c|}{number of scoring events} &  \multicolumn{1}{c}{mean events}\\
 sport & abbrv. & seasons & \multicolumn{1}{|c}{teams} & \multicolumn{1}{|c}{total} & \multicolumn{1}{c}{preprocessed} & \multicolumn{1}{|c}{total}  & \multicolumn{1}{c}{preprocessed} &  \multicolumn{1}{|c}{(preprocessed)} \\ \hline
 Football (college) & CFB & 10, 2000--2009 & 461 & 14,588 & 13,689 & 190,337 & 117,752 & 8.60 \\
 Football (pro)     & NFL & 10, 2000--2009 & 32 & 2,645 & 2,561 & 32,800 & 20,115 & 7.85 \\
 Basketball (pro)   & NBA &  \hspace{0.5em}9, 2002--2010 & 30 & 11,744 & 11,744 & 1,301,408 & 1,096,179 & 93.34\\
 Hockey (pro)       & NHL &  \hspace{0.5em}9, 2000--2009 & 30 & 11,813 & 10,259 & 65,085 & 59,227 & 5.77 \\
\end{tabular}
\end{table*}

\section{Sports datasets}
We use scoring event data\footnote{Data provided by STATS LLC, copyright 2015} from four team sports: college-level American football (CFB, 10 seasons; 2000-2009), professional American football (NFL, 10 seasons; 2000-2009), hockey (NHL, 9 seasons; 2000-2003, 2005-2009)\footnote{The entire 2004 NHL season was canceled due to an extensive lockout over a dispute about player salary caps~\cite{staudohar2005hockey}.} and basketball (NBA, 9 seasons; 2002-2010).  Each dataset consists of the set of scoring events for each game played in the season.  It includes the time the event was scored, the team and player that scored, and its point value.  Table \ref{table:game_stats} gives a summary of these data including the number of teams, games, and individual scoring events. In our analysis and modeling, we discard the timestamps of the events and instead consider only the order in which events appear within a game.

\subsection{Preprocessing}
We extract from the raw event data two sequences to represent each game:\ 
a \textit{point sequence} $\phi$, where $\phi_{i}$ is the point value of scoring event $i$ in the game, and a \textit{team sequence} $\psi$, where $\psi_{i}\in\{r,b\}$ is the identity of the team that won those points. If there are $N_{t}$ events in game $t$, then the corresponding $\phi$ and $\psi$ each contain $N_{t}$ elements, and the lead size at event $i$ is 
\begin{equation}
  L_{i}=\sum_{j=1}^{i}\phi_{j}\delta(\psi_{j},r) - \phi_{j}\delta(\psi_{j},b) \enspace ,
\end{equation}
%
 for team labels $r$ and $b$ (arbitrarily chosen), where $\delta(.,.)$ is the Kronecker delta function and by convention we compute $L$ from team $r$'s perspective.

We begin by removing some games and scoring events.  We remove any events that occur during regulation overtime (0.88\% of all events), because these events follow different scoring processes than events in regular game time~\cite{merritt2014scoring}. Additionally, any games in which only one team scored are removed (6.24\% of all games), as the raw data do not indicate the identity of the non-scoring team. 

Under certain game conditions, multiple scoring events, potentially by different teams, can occur at the same game clock time. For example, in American football, the clock is stopped after a touchdown is scored but the scoring team gets a chance to score a conversion.  If the conversion is unsuccessful, occasionally the opposing team gains control and scores points before the clock is restarted.  Similarly, in basketball, the clock is stopped during free throws after a foul, after which the ball is inbounded (thrown in).  If the ball is inbounded close to the other basket, it is possible to score before a second has elapsed on the clock.  In these cases, the ordering of these events is ambiguous.

Removing these events would alter the running lead size, which is one of the game states of interest. Instead, we merge simultaneous events into a single scoring play that removes the ordering ambiguity while preserving the correct score. If one team scores two simultaneous events $i$ and $i+1$, we merge their values, setting $\phi_{i}=\phi_{i}+\phi_{i+1}$, and removing event $i+1$ from both sequences. If two teams score simultaneously, we merge their values with that of the immediately preceding event in a way that preserves the running lead. Specifically, we set $\phi_{i-1}=\phi_{i-1}\pm|\phi_{i}-\phi_{i+1}|$, where the sign is consistent with the previous assignment of $r$ and $b$ labels to teams, and then remove events $i$ and $i+1$ from both sequences.

\begin{figure}
  \includegraphics[width=\columnwidth]{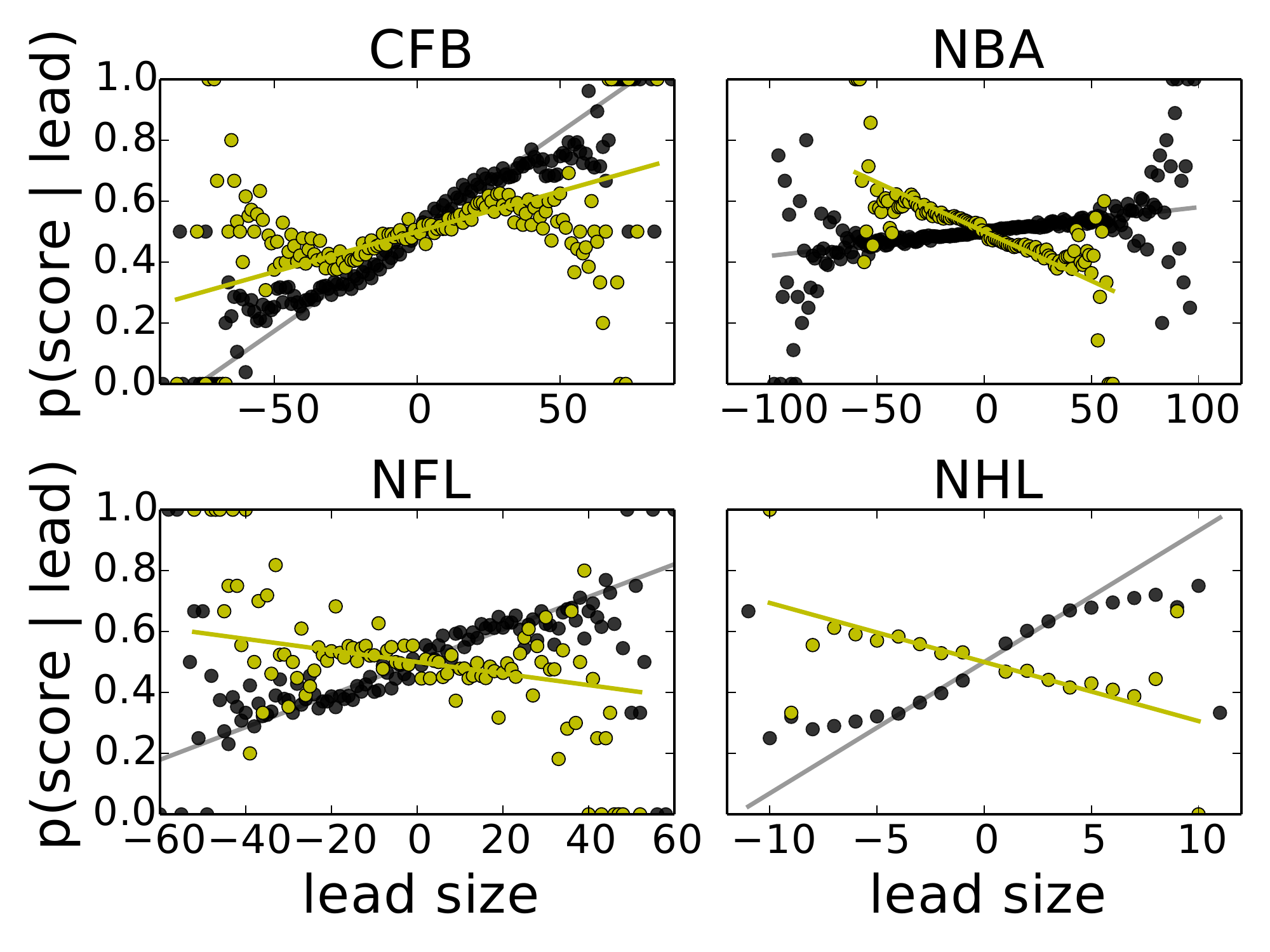}
  \caption{Probability that a team scores next as a function of its lead size, for the observed (\textit{yellow}) and simulated (\textit{black}) patterns, each with a linear least squares fit line. The simulated scoring sequence assumes that the probability of scoring is independent of the game's state.}
  \label{fig:score_lead}
\end{figure} 

\subsection{Scoring and lead size}
\label{sec:scoring:function}
We use these point and team sequences to make an initial investigation of our hypotheses. If the scoring dynamics are truly independent of the game's state, these dynamics will be indistinguishable from an independent Bernoulli process, in which each Bernoulli trial represents a scoring event. We evaluate this model by calculating the empirical probability that a team will score the next event as a function of the current lead size $L$. Recall that we compute $L$ from the perspective of team $r$; thus, if $r$ is leading, then $L$ is positive, while if $r$ is trailing, then $L$ is negative (and vice versa for $b$). This function is thus rotationally symmetric about a lead of $L=0$, where neither team leads, and has the mathematical form of $P(\psi_{i}=r\,|\, L_{i-1})=1-P(\psi_{i}=b\,|\, -\!L_{i-1})$.

We compare the empirical scoring function to one calculated from synthetic team sequences generated according to an independent Bernoulli process, in which we flip a biased coin to determine which team wins each scoring event. The coin's bias is determined by the proportion of scoring events each team wins in that particular game, $\frac{1}{N}\sum_{i=1}^{N}\delta(\psi_{i},r)$ (or for $b$). In this simulation, events are thus independent of the game state (hypothesis 1). We also compute a least-squares regression line for the empirical and for the synthetic data, in which each point is given weight proportional to the number of times the corresponding lead size was observed.  

All of the resulting gradients relating scoring probability to lead size are nonzero (Fig.~\ref{fig:score_lead}), and each Bernoulli process produces a positive gradient. This pattern simply reflects the empirical distribution of biases used to simulate the ensemble of games, with a more positive slope reflecting broader variance in these biases. The variance in the estimated scoring probability increases with lead size simply because progressively fewer games produce leads of that magnitude.

Comparing the observed and simulated scoring functions (Fig.~\ref{fig:score_lead}), we observe a clear contradiction. The gradient and, in particular for NBA, the range of lead sizes generated by the Bernoulli process disagree strongly with those properties observed in the empirical data.   These results suggest that the probability of scoring does indeed depend, somehow, on the game state (hypothesis 2). In subsequent sections, we investigate this dependence using sophisticated probabilistic models to determine how the probability of scoring depends on game state.

\section{Where standard tests fail}
To determine whether scoring events are independent, we now apply a suite of statistical randomization tests, which compare observed sequences to random sequences with similar properties. Specifically, we employ the
\begin{itemize}
\itemsep-1.0pt
  \item serial test (\textit{non-uniformity}),
  \item Wald-Wolfowitz runs test (\textit{anti-restoration}), and
  \item autocorrelation test (\textit{persistence/anti-persistence}),
\end{itemize}
where for each the null hypothesis is that the team sequence $\psi$ is simply a random sequence.

The serial test \cite{knuth1998art} examines bigram frequencies in a sequence and compares them to their expected frequencies under a uniformly random sequence.  For a team sequence with $N$ elements, the observed fractions of bigrams $\{ rr, rb, br, bb \}$ are compared to their expectations of $N/4$.  This test can identify the existence of a bias within each game, i.e., if one team is systematically more likely to score than another.

The Wald-Wolfowitz runs test~\cite{wald1940test} examines the observed number of runs in a sequence, i.e., substrings of $\psi$ for which each element is the same (either $r$ or $b$), which allows us to identify either positive momentum or anti-restorative effects in within-game scoring. 
We reject the null hypothesis that $\psi$ is random if the observed number of runs is significantly below its expected value.  Previously, this test has been used to detect winning streaks in sequences of games~\cite{vergin2000winning}.

The autocorrelation test measures the correlation of a sequence with itself, shifted by one element, which allows us to identify periodic dynamics that occur as a result of anti-persistence. Here, we reject the null hypothesis that $\psi$ contains no dependence between values if the autocorrelation is significantly higher or lower than zero.

\begin{figure}
  \includegraphics[width=\columnwidth]{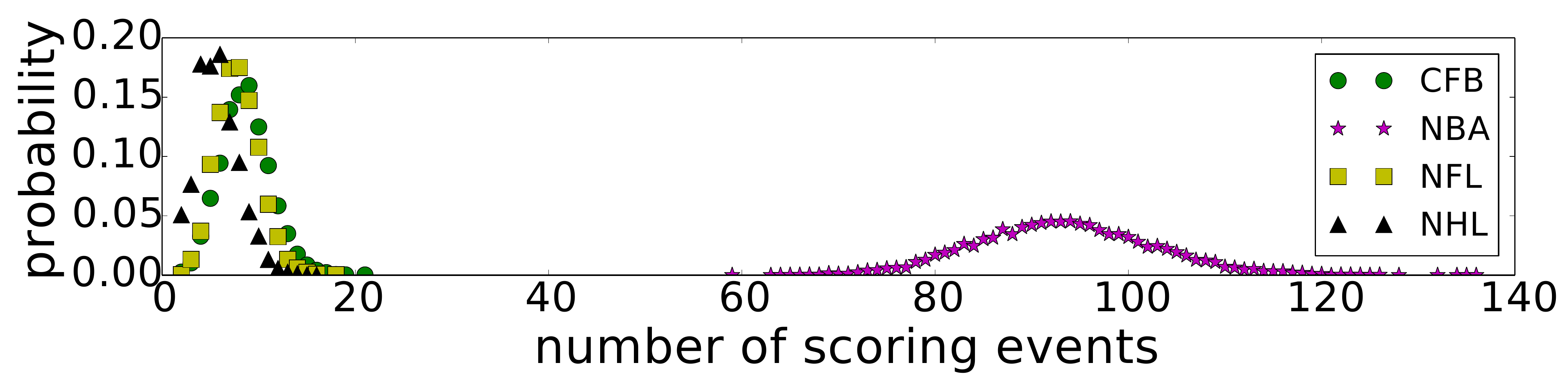}\\
  \includegraphics[width=\columnwidth]{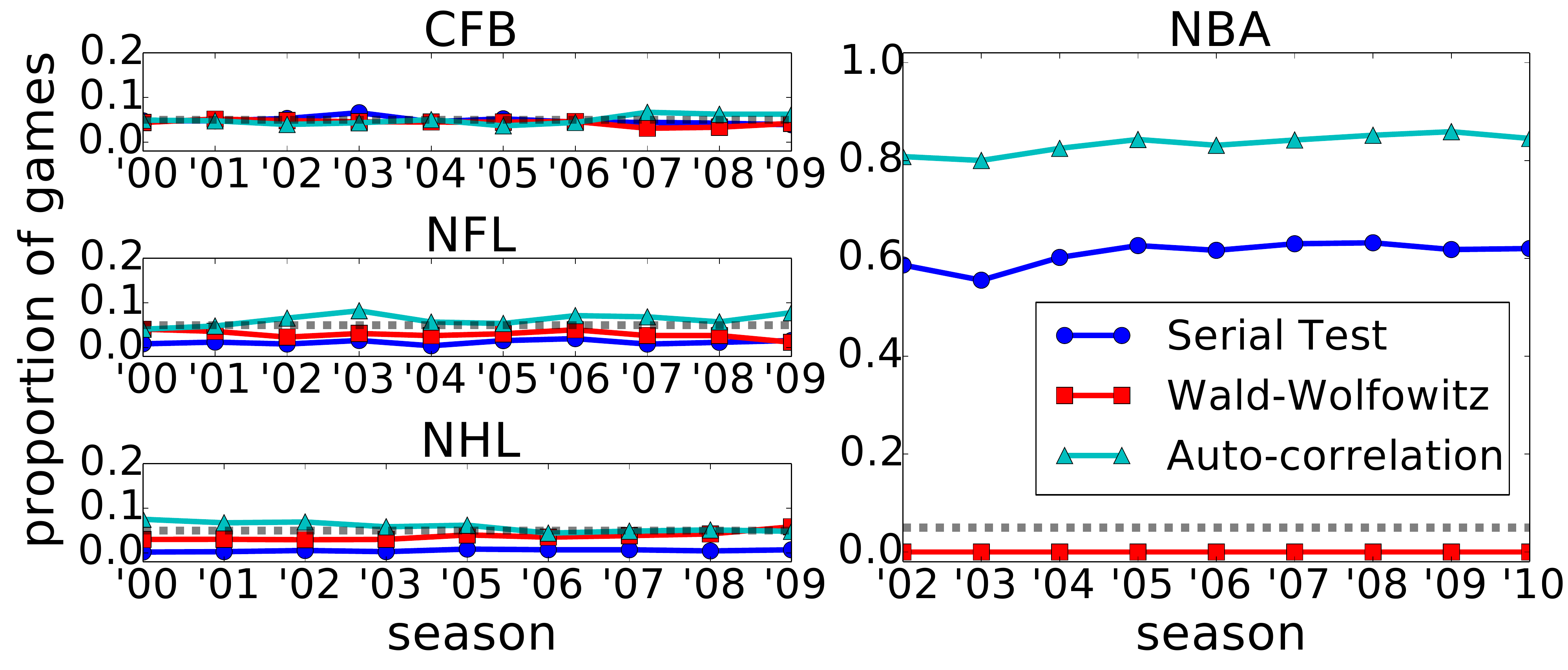}
  \caption{(\textit{top}) Probability distributions for the number of scoring events in a game, and (\textit{bottom}) the randomization test results for each sport, by season, versus a false positive rate of $\alpha=0.05$ (dashed line). The team sequences of each game are tested independently and we plot the proportion of games that reject the null hypothesis that the sequences are random.  Because CFB, NFL and NHL typically have a small number of events per game (upper panel), the null hypothesis is difficult to reject.}
  \label{fig:skids}
\end{figure}

We apply each of these three tests to each of our four data sets, and compare the results against a false positive rate of $\alpha=0.05$ (Fig.~\ref{fig:skids}). We also consider each season separately so as to reveal non-stationarities.
Basketball, unlike the other sports, produces a large proportion of rejections for the serial and autocorrelation tests, which reflects the known anti-persistence pattern in basketball scoring~\cite{gabel2012random}.

On the other hand, for all sports except basketball, each of these tests rejects the null hypothesis at close to or below the chosen false positive rate, a finding consistent with each of these sequences being random. However, this interpretation is problematic. The serial test makes the very strict assumption that each sequence is drawn from a uniform random distribution, i.e., each is generated by flipping a fair coin several times. A face-value interpretation thus implies that all teams have an equal chance of winning each game---a highly unlikely situation---and it predicts that the scoring function from Section~\ref{sec:scoring:function} should be independent of lead size, which contradicts the observed pattern (Fig.~\ref{fig:score_lead}).

In fact, however, there is no contradiction:\ the $\psi$ sequences are simply too short (Fig.~\ref{fig:score_lead}) for these tests to reliably distinguish random from non-random sequences when we assume they are generated independently, i.e., the tests have low statistical power. The one exception is basketball, whose sequences typically contain 90 or so events, while those for American football or hockey typically contain less than 10.

In the following sections, we show how to circumvent the low statistical power of these tests by exploiting the fact that team sequences are not, in fact, independent of each other. Instead, each season's sequences are generated by repeatedly selecting pairs from a finite and fixed population of teams. This process induces substantial correlations across games that we can capture by modeling the latent skills of each team within a given season.

\section{Skill-based scoring dynamics}
Toward this end, we develop a series of models of increasing complexity based on specific underlying mechanisms for sports scoring dynamics, including independence, restoration, and anti-persistence.  Each of these models represents team skill as a latent variable. We assume that team skill is fixed over the course of any particular season~\cite{glickman1998state}, which reflects the relatively stable team rosters and coaching staffs, and low injuries rates in these sports.  
Furthermore, modeling each season separately allows us to run multiple tests for each sport---one for each season---and allows our models to capture real changes in team skill across seasons~\cite{glickman1998state}.

Each of our models generates a team sequence $\psi$ by extending the popular Bradley-Terry (BT) model~\cite{bradley1952rank} to generate individual scoring events within a game. Traditionally, the BT model is used to estimate unobserved (latent) team skills from the observed outcomes of many games among pairs of teams. The probability that team $r$ wins in a match against team $b$ is given by the skill of $r$ relative to $b$:
\begin{equation}
\label{eq:BT}
  P(r~\textrm{wins against}~b) = d_{rb} = \frac{\pi_r}{\pi_r + \pi_b} \enspace ,
\end{equation} 
where $\pi_r,\pi_{b} \in [0,1]$ is the latent skill for team $r$.

\subsection{Independent model}
When scoring events within a game are independent, their generation is equivalent to a simple Bernoulli process with a game-specific bias. 
This is equivalent to an ``independent model'' that applies the game-level BT model of Eq.~\eqref{eq:BT} to each of the individual scoring events within a game, yielding
\begin{equation}
  P(\psi_{i}=r) = d_{rb} \enspace .
\end{equation}
This represents our first model, which can capture variability in a team sequence caused by differences in team skill parameters, but not other sources of variability.

\subsection{Restorative models}
\label{sec:restorative}
Real scoring functions (Fig.~\ref{fig:score_lead}) produce a range of gradients. However, the independent model can only produce positive slopes. To capture a wider variety of scoring function shapes, and in particular a negative slope or ``restorative'' pattern, we extend the independent model by allowing each team's skill to explicitly covary with its lead. 
Such a relationship could arise for psychological reasons, e.g., a winning team ``loses steam'' or a losing team gains motivation~\cite{berger2011can}, or for strategic reasons, e.g., substituting out or in the more skilled players while in the lead in order to conserve their energy, avoid injury, or create momentum~\cite{merritt2014scoring}.

Our restorative model augments the independent model with an explicit per-team ``restorative force'' parameter $\gamma_{r}$, which modifies team $r$'s strength in response to the current lead size from its perspective $\ell_{r}$ and captures the fact that different teams may have different behaviors in response to how far ahead or behind they are. When $\gamma_{r}<0$, team $r$ exhibits a restorative pattern, with skill being proportional to $-\ell_{r}$. When $\gamma_{r}>0$, team $r$ exhibits an anti-restorative or momentum pattern, with skill being proportional to $\ell_{r}$.

The probability that team $r$ scores against $b$ is given by
\begin{equation}
  P(\psi_{i}=r) = d_{rb} + \ell_{ir} c_{rb} \enspace,
  \label{eq:linear_momentum}
\end{equation}
where $\ell_{ir}$ is $r$'s lead size just before event $i$ and 
\begin{equation}
  c_{rb} = \gamma_r + \gamma_b = c_{br} \enspace.
\end{equation}

A game as a whole exhibits a restorative pattern whenever $c_{rb}<0$. This occurs either when both teams exhibit a restorative pattern themselves ($\gamma_{r}<0$ and $\gamma_{b}<0$) or when one team's restorative force is stronger than the other team's anti-restorative force ($\gamma_{r}<0$, $\gamma_{b}>0$, and $|\gamma_{r}|>|\gamma_{b}|$).

The additional term in Eq.~\eqref{eq:linear_momentum} relative to the independent model means this model's scoring function is no longer bounded on the $[0,1]$ interval. We correct this behavior by using a sigmoid function of the form $\sigma (x) = \left(1+\textrm{e}^{-x}\right)^{-1}$ to provide a smooth and continuous approximation of the misspecified linear function.

To make this approximation, we change variables so that a logistic curve most closely approximates the linear equation, which occurs when we match the gradients at the point of symmetry at $P(\psi_{i}=r) = 1/2$. Setting the derivative $\sigma'$ equal to $c_{rb}$, we find
\begin{equation}
  \sigma' (m_{rb}\ell_{ir} + v_{rb}) = \frac{m_{rb}\textrm{e}^{m_{rb}\ell_{ir} + v_{rb}}}{(\textrm{e}^{m_{rb}\ell_{ir}} + \textrm{e}^{v_{rb}})^2} = c_{rb} \enspace ,
  \label{eq:sigma_prime}
\end{equation}
We then solve for when the logistic function equals $1/2$, yielding
\begin{equation}
  \sigma (m_{rb}\ell_{ir} + v_{rb}) = \frac{1}{1+\textrm{e}^{-(m_{rb}\ell_{ir} + v_{rb})}} = 1/2 \enspace . \label{eq:sigma} 
\end{equation}

\begin{figure}
\centering
  \includegraphics[width=\columnwidth]{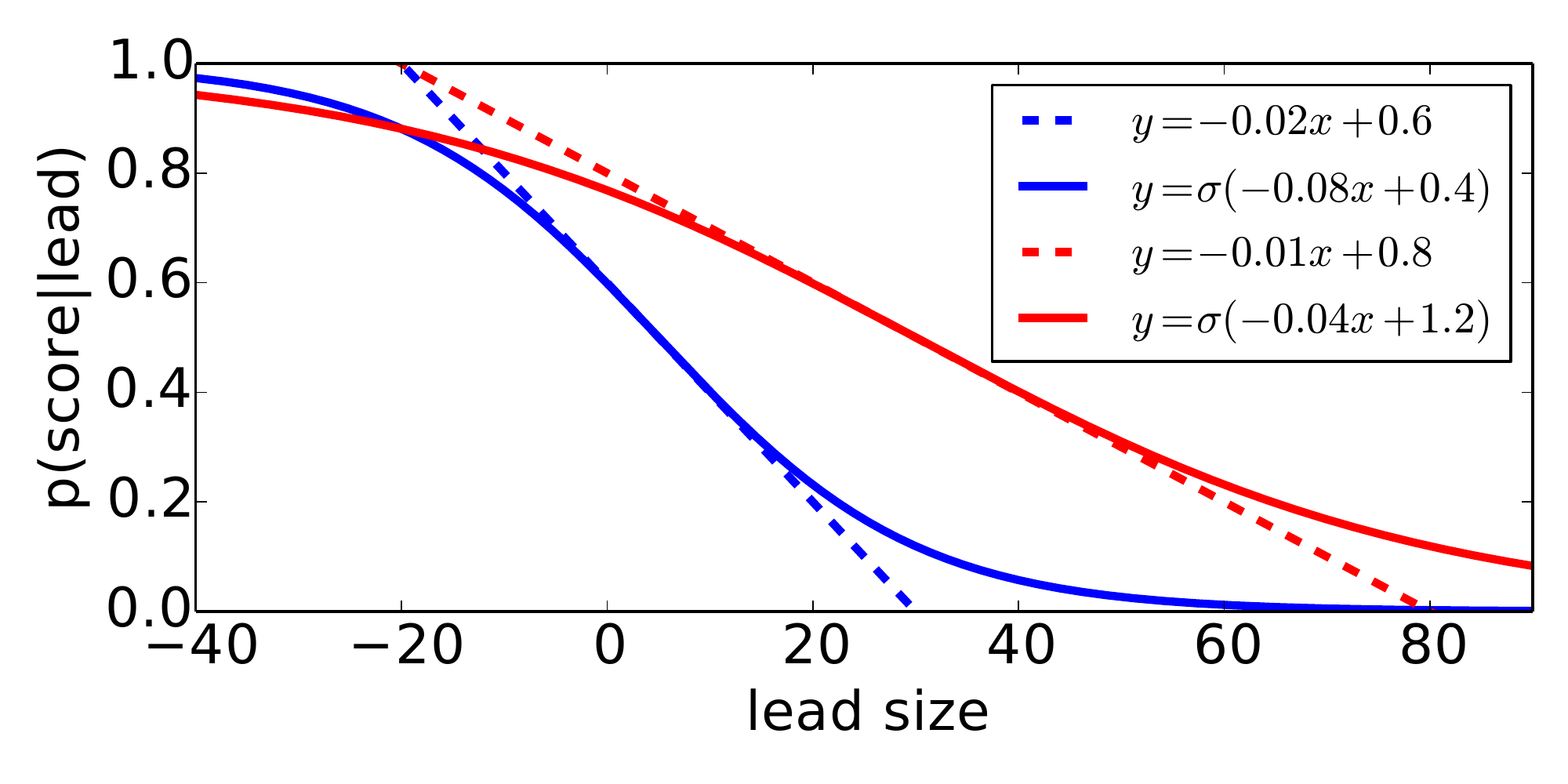}
  \caption{Two examples of linear functions matched to logistic functions using the change of variables in Eqs. \eqref{eq:v} and \eqref{eq:m}.}
  \label{fig:chg_var}
\end{figure}

Finally, in solving Eqs.~\eqref{eq:sigma_prime} and~\eqref{eq:sigma} we obtain the following transformation of variables:
\begin{align}
  v_{rb} &= -4\left(1/2-d_{rb}\right) \label{eq:v}\\
  m_{rb} &= 4\,c_{rb} \enspace ,  \label{eq:m}
\end{align}
where $m_{rb}$ and $v_{rb}$ are the variables used in the logistic function such that $c_{rb}$ and $d_{rb}$ retain their linear interpretation and are thus comparable to the skill variables in the independent scoring model.  Figure~\ref{fig:chg_var} shows examples of two linear functions and their corresponding logistic approximations.

\begin{table*}
\centering
\caption{Log-likelihoods on held-out data for NBA games.}
\label{table:NBA_LL}
\begin{tabular*}{\textwidth}{|@{\extracolsep{\fill} }l|rrrrrrrrr|}
\hline
   & 2002 & 2003 & 2004 & 2005 & 2006 & 2007 & 2008 & 2009 & 2010 \\ \hline
    Independent & -80849 & -78814 & -84698 & -84744 & -84795 & -86070 & -85727 & -86314 & -85114 \\
    Restorative & -80573 & -78506 & -84361 & -84404 & -84469 & -85777 & -85444 & -86005 & -84704 \\
    Independent anti-persistent & \color{blue!50!black} -75655 & \color{blue!50!black}-73823 & \color{blue!50!black}-79151 & \color{blue!50!black}-78841 & \color{blue!50!black}-79088 & \color{blue!50!black}-80174 & \color{blue!50!black}-79841 & \color{blue!50!black}-80513 & \color{blue!50!black}-79386  \\
    Restorative anti-persistent & \textbf{\color{blue!50!black}-75627} & \textbf{\color{blue!50!black}-73777} & \textbf{\color{blue!50!black}-79097} & \textbf{\color{blue!50!black}-78796} & \textbf{\color{blue!50!black}-79040} & \textbf{\color{blue!50!black}-80141} & \textbf{\color{blue!50!black}-79812} & \textbf{\color{blue!50!black}-80465} & \textbf{\color{blue!50!black}-79297}  \\
    \hline
\end{tabular*}

\vspace{-0.5mm}
\caption{Log-likelihoods on held-out data for NFL games.}
\label{table:NFL_LL}
\begin{tabular*}{\textwidth}{|@{\extracolsep{\fill} }l|rrrrrrrrrr|}
\hline
  & 2000 & 2001 & 2002 & 2003 & 2004 & 2005 & 2006 & 2007 & 2008 & 2009 \\ \hline
  Independent & \color{blue!50!black}-1286 & \color{blue!50!black}-1307 & \color{blue!50!black}-1408 & \color{blue!50!black}-1372 & \color{blue!50!black}-1403 & \textbf{\color{blue!50!black}-1373} & \textbf{\color{blue!50!black}-1369} & \color{blue!50!black}-1433 & \color{blue!50!black}-1484 & \color{blue!50!black}-1395 \\
  Restorative & -1324 & -1347 & -1450 & -1402 & -1451 & -1422 & -1424 & -1466 & -1530 & -1432 \\
  Independent anti-persistent & \textbf{\color{blue!50!black}-1278} & \textbf{\color{blue!50!black}-1290} & \textbf{\color{blue!50!black}-1401} & \textbf{\color{blue!50!black}-1361} & \textbf{\color{blue!50!black}-1392} & \color{blue!50!black}-1378 & \color{blue!50!black}-1372 & \textbf{\color{blue!50!black}-1425} & \textbf{\color{blue!50!black}-1473} & \textbf{\color{blue!50!black}-1387} \\
  Restorative anti-persistent & -1322 & -1337 & -1450 & -1496 & -1448 & -1427 & -1434 & -1470 & -1520 & -1426 \\
  \hline
\end{tabular*}

\vspace{-0.5mm}
\caption{Log-likelihoods on held-out data for CFB games.}
\label{table:CFB_LL}
\begin{tabular*}{\textwidth}{|@{\extracolsep{\fill} }l|rrrrrrrrrr|}
\hline
  & 2000 & 2001 & 2002 & 2003 & 2004 & 2005 & 2006 & 2007 & 2008 & 2009 \\ \hline
  Independent & \color{blue!50!black}-7487 & \textbf{\color{blue!50!black}-7575} & \textbf{\color{blue!50!black}-8098} & \textbf{\color{blue!50!black}-8105} & \textbf{\color{blue!50!black}-7675} & \textbf{\color{blue!50!black}-7708} & \textbf{\color{blue!50!black}-7265} & \textbf{\color{blue!50!black}-8673} & \textbf{\color{blue!50!black}-8435} & \color{blue!50!black}-8097 \\
  Restorative & -8114 & -8182 & -8689 & -8656 & -8268 & -8176 & -7884 & -9334 & -9065 & -8777 \\
  Independent anti-persistent & \textbf{\color{blue!50!black}-7486} & \color{blue!50!black}-7643 & \color{blue!50!black}-8142 & \color{blue!50!black}-8201 & \color{blue!50!black}-7741 & \color{blue!50!black}-7759 & \color{blue!50!black}-7328 & \color{blue!50!black}-8678 & \color{blue!50!black}-8458 & \textbf{\color{blue!50!black}-8078} \\
  Restorative anti-persistent & -8011 & -8113 & -8625 & -8586 & -8198 & -8110 & -7781 & -9214 & -8880 & -8630 \\
  \hline
\end{tabular*}

\vspace{-0.5mm}
\caption{Log-likelihoods on held-out data for NHL games.}
\label{table:NHL_LL}
\begin{tabular*}{\textwidth}{|@{\extracolsep{\fill} }l|rrrrrrrrr|}
\hline
  & 2000 & 2001 & 2002 & 2003 & 2005 & 2006 & 2007 & 2008 & 2009 \\ \hline
  Independent & \color{blue!50!black}-4432 & \color{blue!50!black}-4238 & \color{blue!50!black}-4300 & -4078 & \color{blue!50!black}-5026 & -4712 & \color{blue!50!black}-4504 & \textbf{\color{blue!50!black}-4755} & \textbf{\color{blue!50!black}-4655} \\
  Restorative & -4432 & -4238 & -4313 & \textbf{\color{blue!50!black}-4056} & -5031 & \textbf{\color{blue!50!black}-4695} & -4511 & -4761 & \color{blue!50!black}-4663 \\
  Independent anti-persistent & \textbf{\color{blue!50!black}-4420} & \textbf{\color{blue!50!black}-4237} & \textbf{\color{blue!50!black}-4287} & \color{blue!50!black}-4068 & \textbf{\color{blue!50!black}-5020} & \color{blue!50!black}-4706 & \textbf{\color{blue!50!black}-4497} & \color{blue!50!black}-4761 & -4668 \\
  Restorative anti-persistent & -4449 & -4254 & -4318 & -4090 & -5045 & -4721 & -4521 & -4787 & -4687 \\
  \hline
\end{tabular*}
\end{table*}

\subsection{Anti-persistence models}
In many sports, we observe an \textit{anti-persistent} pattern in the team sequences, in which the probability that $r$ scores next depends on which team scored last, i.e., $P(\psi_{i+1}=r\,|\,\psi_{i})$. For example, for NBA team sequences, the rate of $rr$ and $bb$ bigrams is only 0.35, indicating strong anti-persistence. (The rates for CFB, NFL, and NHL are 0.45, 0.44, and 0.49, respectively.) Such an anti-persistence pattern can occur when teams have different degrees of skill at defensive and offensive play, e.g., when both teams have offenses that are relatively stronger than the opposing team's defense. 

To capture these effects, we extend the independent model so that each team has an offensive skill parameter $\pi^{\rm{off}}$ and a defensive parameter $\pi^{\rm{def}}$. For sports like American football and basketball, ball possession (offensive play) typically alternates after a scoring event. We model this game rule by applying a team's defensive skill immediately after it scores and its offensive skill after the other team scores. Under this independent anti-persistent model, the probability of scoring event $i$ is 
\begin{equation}
P(\psi_{i} \!=\! r \,|\,  \psi_{i-1}) =
\left\{ 
\begin{array}{ll}
\!\! \pi_r^{\rm{def}} \left/ \left(\pi_r^{\rm{def}} \!+\! \pi_b^{\rm{off}}\right) \right.  & \textrm{if}~\psi_{i-1}=r  \vspace{0.5em} \\
\!\! \pi_r^{\rm{off}} \left/ \left(\pi_r^{\rm{off}} \!+\! \pi_b^{\rm{def}}\right) \right. & \textrm{if}~\psi_{i-1}=b \\
\end{array}
\right.  .
\end{equation}
Finally, we obtain a fourth model by combining the restorative model with the anti-persistent model. 

\section{Modeling scoring dynamics}
\label{sec:scoring}
We fit the (i)~independent, (ii)~restorative, (iii)~independent anti-persistent, and (iv)~restorative anti-persistent models to the team sequences within a given season of each sport, using Markov chain Monte Carlo to estimate each model's parameters.  For each, we assess model goodness-of-fit by calculating the held out likelihood for each model under a 10-fold cross validation. Furthermore, we follow this procedure for each season of each sport separately, the results of which are given in Tables~\ref{table:NBA_LL}--\ref{table:NHL_LL}. By treating seasons independently, we obtain multiple model assessments within each sport while controlling for within season variability. 
For each season, we highlight the two highest scores in blue and the highest score in bold.

In basketball (NBA), we find that the restorative anti-persistent model consistently provides the best fit across all seasons (Table~\ref{table:NBA_LL}), with the second best model being the independent anti-persistent model. These results indicate a strong role for both restoration and anti-persistence in driving basketball scoring dynamics. Previous analysis of basketball scoring using random walk theory came to similar conclusions~\cite{gabel2012random}.

American football (NFL and CFB) shows a different result, with both types of independent model being heavily favored over both types of restorative model (Tables \ref{table:NFL_LL} and \ref{table:CFB_LL}). The poor fit here of the restorative models indicates that the competitive processes that produce a restorative force in basketball are largely absent in American football. This difference may be related to the much greater scoring rate in basketball relative to American football (Fig.~\ref{fig:skids}):\ an increased scoring rate lowers the marginal value of each scoring event relative to the game outcome (who wins), and low value interactions in other systems are associated with restorative forces~\cite{durham1998rich,hartley2007handbook}.

Furthermore, the anti-persistent model for NFL is favored in 8 of 10 seasons over the independent model, while in CFB, it is favored in only 2 of 10 seasons. That is, anti-persistence appears to play a stronger role in NFL games than in CFB games. In fact, CFB is the only sport to strongly favor the independent model, a result that agrees with the our previous simulation results (Fig.~\ref{fig:score_lead}), which showed that the trivial independent model produced the smallest disagreement for CFB between real and simulated scoring function gradients among the four sports.

The results for hockey (NHL) are less clear cut (Table~\ref{table:NHL_LL}). In 8 out of 9 seasons, the independent anti-persistent model is either the best or second best model, and the independent model is best or second best in 7 out of 9. On the other hand, the simple restorative model wins for 2 seasons, and is second best for one. (The restorative anti-persistent model is a poor fit for all hockey seasons.) We note, however, that the log-likelihoods among these three models are all very close, indicating that each performs about as well as the others for these data. Given that NHL is also the one sport among the four that is not anti-persistent by design (possession is determined by a ``faceoff'' after each goal) and that its scoring function has a negative gradient, we tentatively conclude that the restorative model is better.

\begin{figure}
  \includegraphics[width=\columnwidth]{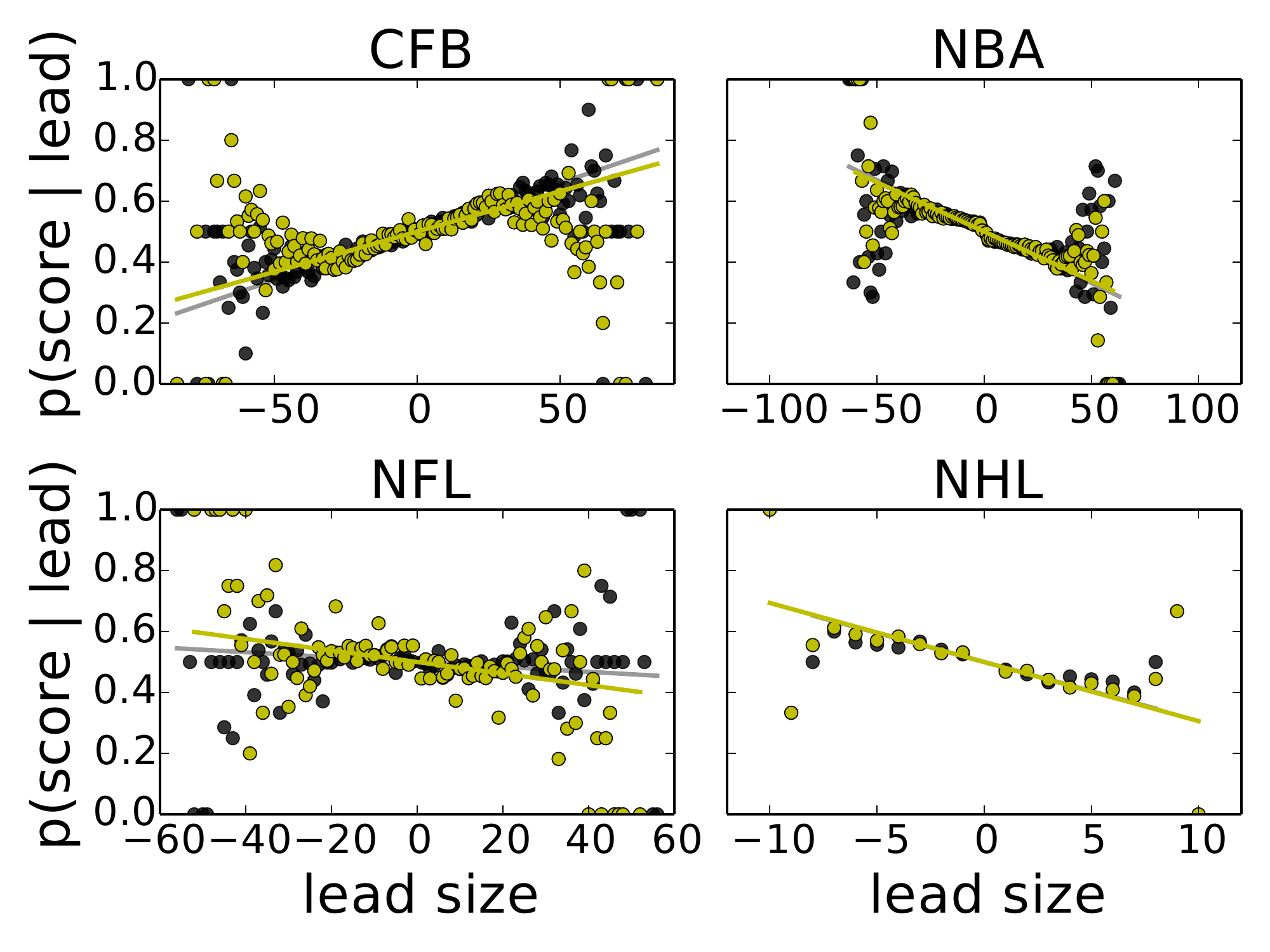}
  \caption{Probability that a team scores next as a function of its lead size, for the observed (\textit{yellow}) and simulated (\textit{black}) patterns, each with a linear least squares fit line. Each simulation uses the best overall skill model for that sport to generate synthetic point and team sequences.}
  \label{fig:score_lead2}
\end{figure} 

Across seasons, the best overall models appear to be CFB:\ independent; NFL:\ independent anti-persistent; NBA:\ restorative anti-persistent; and NHL:\ restorative. We check these models by performing a semi-parametric bootstrap, generating synthetic $\phi$ and $\psi$ sequences of the same number and lengths as observed empirically in each season, and comparing the simulated and empirical scoring functions. That is, we repeat the assessment of Figure~\ref{fig:score_lead}, but now using models that can capture dependence across sequences.
The results show that our skill-based models are a dramatic improvement over simulating each game independently (Fig.~\ref{fig:score_lead2}), agreeing closely with the empirical scoring patterns in both the gradient and range of lead sizes.

\section{Predicting Outcomes}
We now apply our models to two online prediction tasks in each of the sports: \textit{Who will score next?} and \textit{Who will win?} For both tasks, we let our models observe the point and team sequences of the first $T$ games in a particular season. We then use these models to predict for each unobserved game in that season (i)~the team sequence values $\psi_{i}$ for $1\leq i \leq N$, and (ii)~the identity of the winning team, when each model is allowed to observe the game states $(\psi_{j},\phi_{j})$ for $1\leq j < i$. In the second task, all models predict point values $\phi_{i}$ as the mean value $\langle \phi \rangle$ averaged over all events in the season. We compare our predictions to those of three baseline models. 

The first baseline is a na\"{\i}ve \textit{leading} model, which assumes that the team currently in the lead is the stronger team and thus more likely to both score next and win the game. Specifically, it predicts that team holding the lead at event $i$ will win the next event, i.e., it predicts $\psi_{i+1}=r$ if $L>0$ and $\psi_{i+1}=b$ if $L<0$, and will also win the game. If $L=0$, the model flips a fair coin for $r$ and $b$.

The second baseline is the standard \textit{Bradley-Terry} model in which we infer latent team skills $\pi$ from the win-loss records among teams in the observed games. This model is simpler than our independent model, which infers team skills using team sequences $\{\psi\}$ of the observed games.

The third baseline is a simple \textit{first order Markov} model.  It predicts that the next team to score will either be the same or different than the team that scored last according to the empirical bigram frequencies $\{rr,bb,rb,br\}$ observed in the first $T$ games of the season.  Formally, it predicts that a team will score again given it scored last time as 
\begin{equation}
  P(\psi_{i+1}\!\!=\!\psi_{i}) = \left. \left(\sum_{t=1}^T{\sum_{i=1}^{N_t-1}{\delta(\psi_{i+1},\psi_{i})}}\right) \middle/ \left(\sum_{t=1}^T{N_t-1}\right) \right. \enspace .
\end{equation}

For both prediction tasks, we assess prediction accuracy via AUC statistic, which gives the probability that a randomly selected true positive is ranked above a randomly selected false positive. The AUC is a statistically principled measure for binary classification tasks like ours where the cost of an error is the same in either direction (since team labels, $r$ or $b$, are arbitrary).

\begin{figure}
\centering
  \includegraphics[width=\columnwidth]{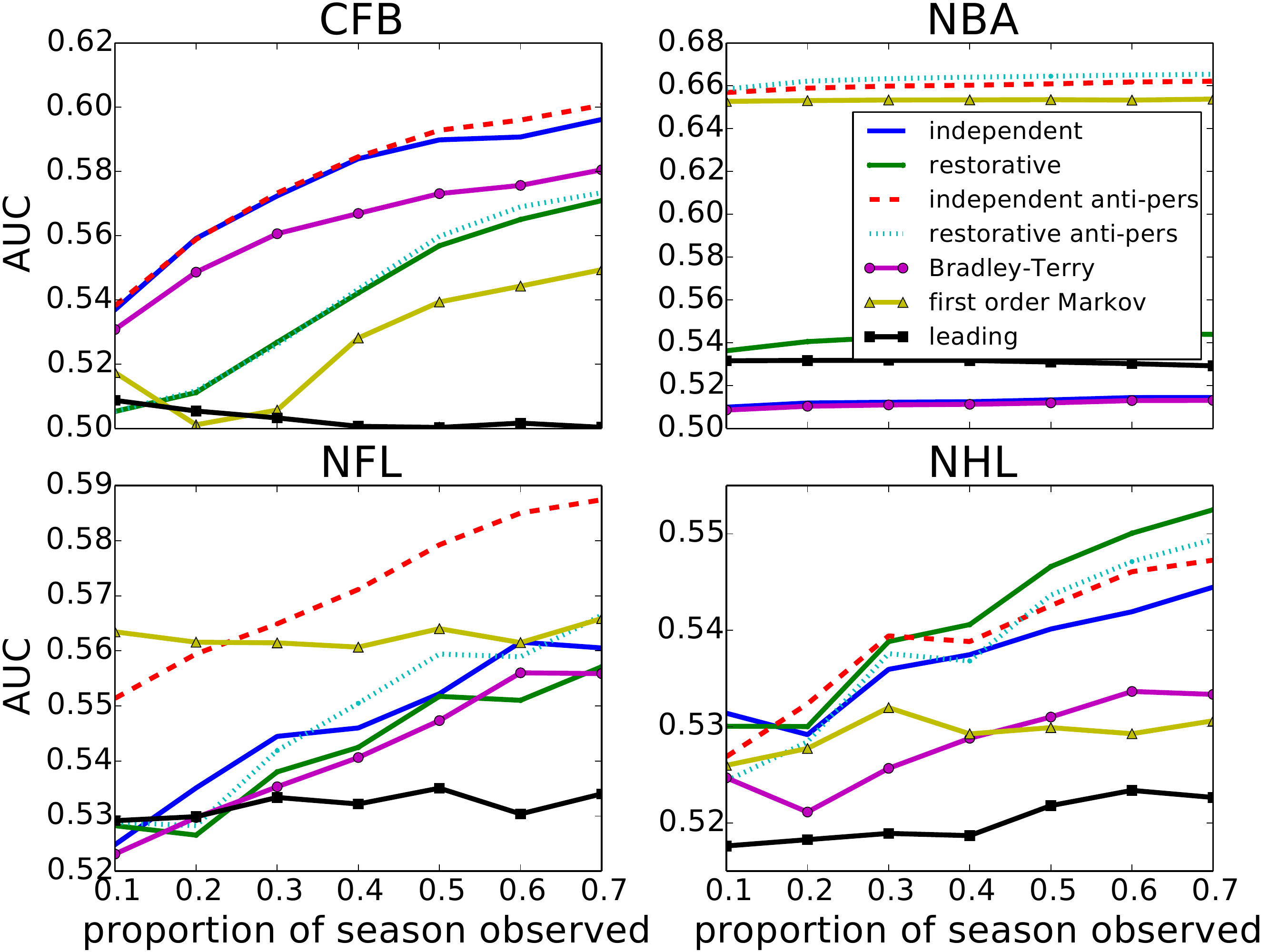}
  \caption{Probability of accurately predicting which team will score next (AUC), when models observe different fractions of a season.  Based on 95\% confidence intervals, our best model performs significantly better than the baseline models for CFB and NBA, and after observing at least half of the season for NFL and NHL. }
  \label{fig:predictnext}
\end{figure}

\subsection{Who will score next?}
In the first task, we aim to predict which team will score event $i$, for each $1 \leq i \leq N$, given the sequence of preceding game states $(\phi_{j},\psi_{j})$ for $1\leq j < i$. For this online prediction task, we learn each model's parameters from the first $T$ games in a season and then make predictions across all unobserved games within a season and calculate the AUC for all predictions across all seasons to obtain a single score. Each model observes at least 10\% of a season, which ensures that every team has played at least a few times.

The results show that the overall best models identified in the previous section also tend to be the best predictors at who will score next (Fig.~\ref{fig:predictnext}), although some alternative models also perform well. For instance, the best model for NFL games early in the season is the first order Markov model; however, the best NFL model beats this baseline after about 30\% of a season is observed. Similarly, the first order Markov model performs almost as well as the best skill model in predicting who will score next in NBA games, by capturing the known anti-persistence pattern in that sport. One of the worst models across all four sports is the leading baseline, which often performs only slightly better than chance.

\subsection{Who will win?}
Predicting who will win a game requires extrapolating the point and team sequences to determine the game's final outcome. We simplify this task slightly by assuming that the number of scoring events $N$ in each game is known. We then allow the models to learn their parameters from the first 30\% of each season (other choices lead to qualitatively similar results as those reported here). For each game in the remainder of a season, the models predict the identity of the winning team when they are allowed to observe a progressively greater fraction of game states $(\phi_{i},\psi_{i})$ for $0.1\leq i/N \leq 0.9$---as if each model were watching the game unfold in real time.

\begin{figure}
\centering
  \includegraphics[width=\columnwidth]{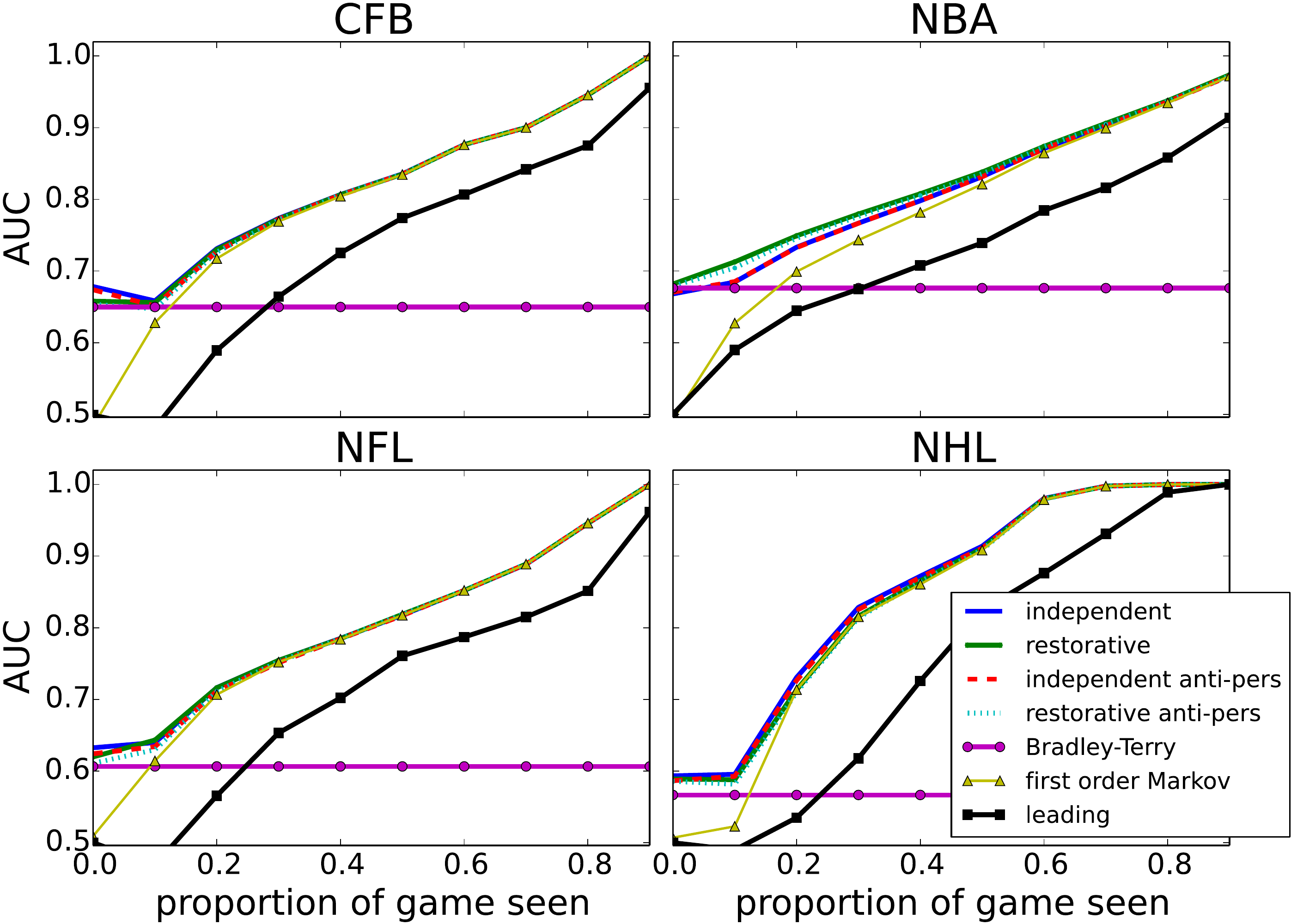}
  \caption{AUC scores for predicting which team will win given the current state of the game.}
  \label{fig:predictwin}
\end{figure}

The results show that the overall best model for each sport both consistently outperforms the baselines and also correctly predicts the winner with at least 80\% accuracy at a game's halftime (Fig.~\ref{fig:predictwin}).

The relatively poorer performance of the ``leading'' baseline model illustrates that this prediction task is non-trivial---who is leading at a given moment is not as predictive of who wins as knowing something about team skills and scoring dynamics. On the other hand, the Bradley-Terry baseline performs comparably well very early in the game, but is quickly beaten because it cannot learn from the real-time evolution of a game.

For this task, most of our skill-based models make very similar predictions and the first order Markov model also performs well. Although the distributions of final lead sizes may be different, the means are very close, and the individual predictions across models correlate strongly. The greatest difference occurs at the start of the game. In particular, the first order Markov model performs much worse than the skill-based models at the beginning because it has no information about the heterogeneity of team scoring abilities.  As the game progresses the predictions tend to converge. This occurs because these models all make predictions based on random walks on a binary sequence $\{r,b\}$, the difference being in how they model the transition probabilities. Later in the game we extrapolate less and so the differences between models become less pronounced. 

\section{Team skill evolves over time}
A useful feature of our probabilistic models is the interpretability of their parameters, which are meaningful measures of team skill here. By learning these parameters independently for each season in each sport, we can investigate how team skills have evolved over time.

\begin{figure}
\centering
  \includegraphics[width=.49\columnwidth]{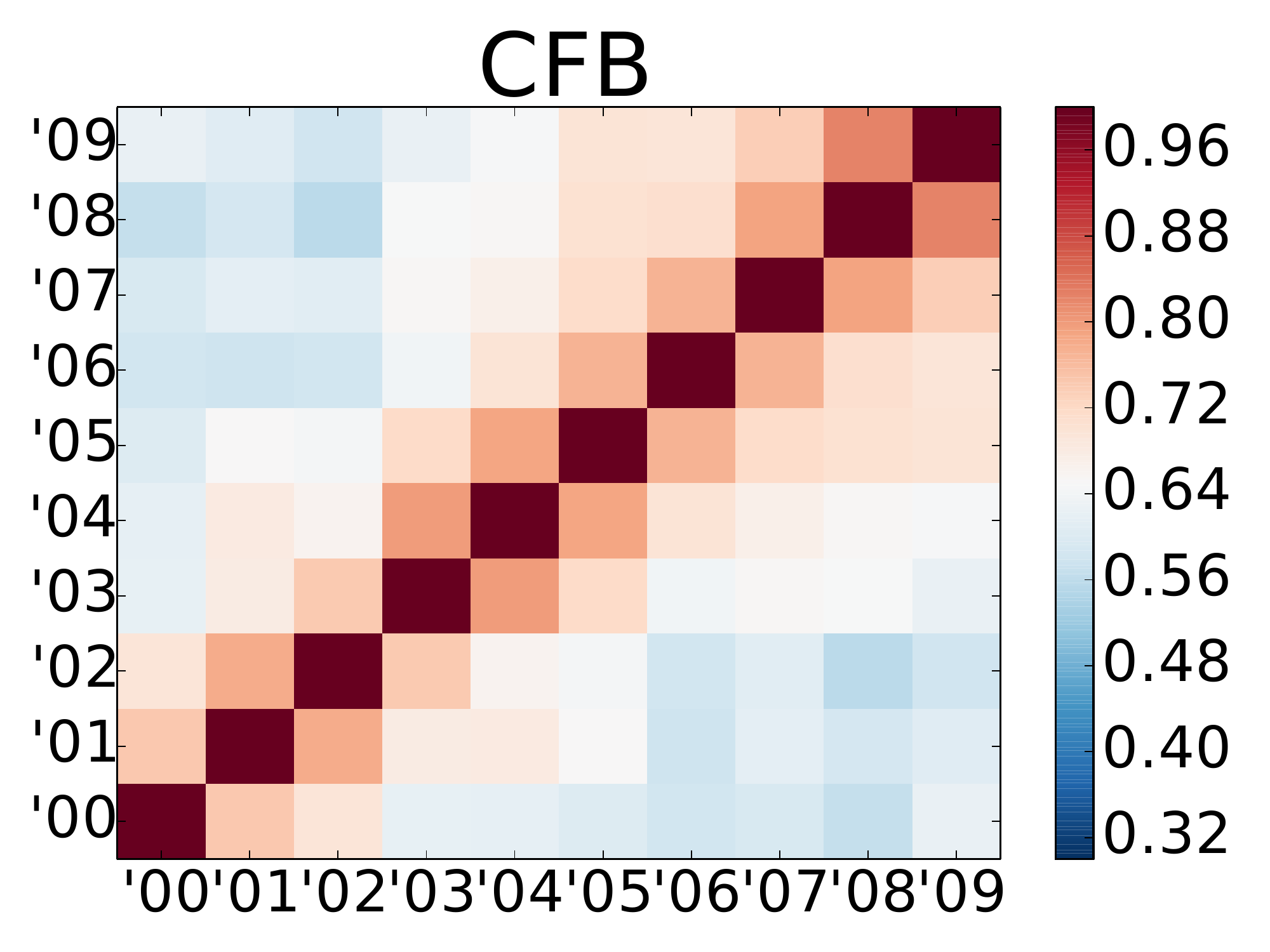}
  \includegraphics[width=.49\columnwidth]{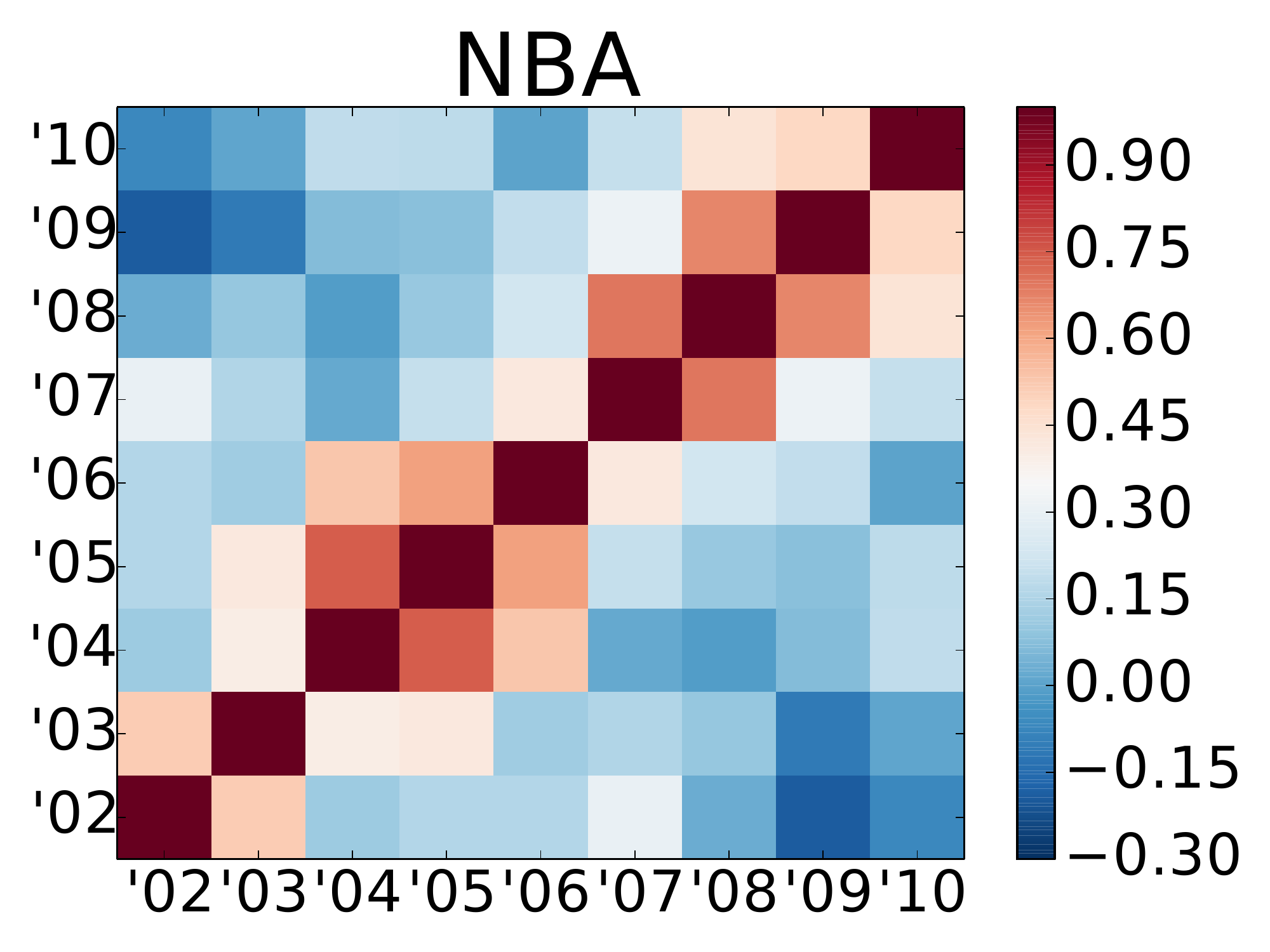} \\
  \includegraphics[width=.49\columnwidth]{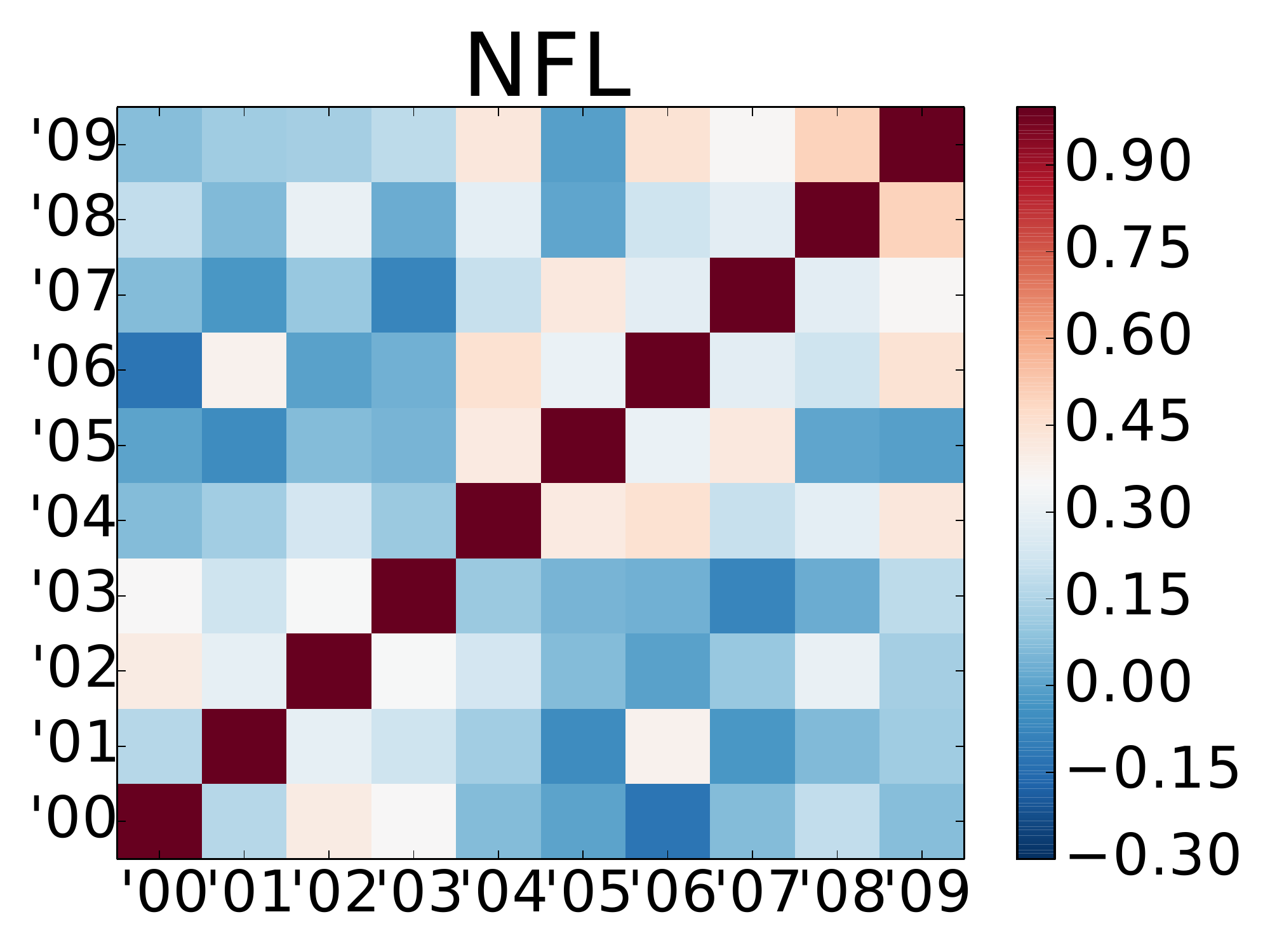}
  \includegraphics[width=.49\columnwidth]{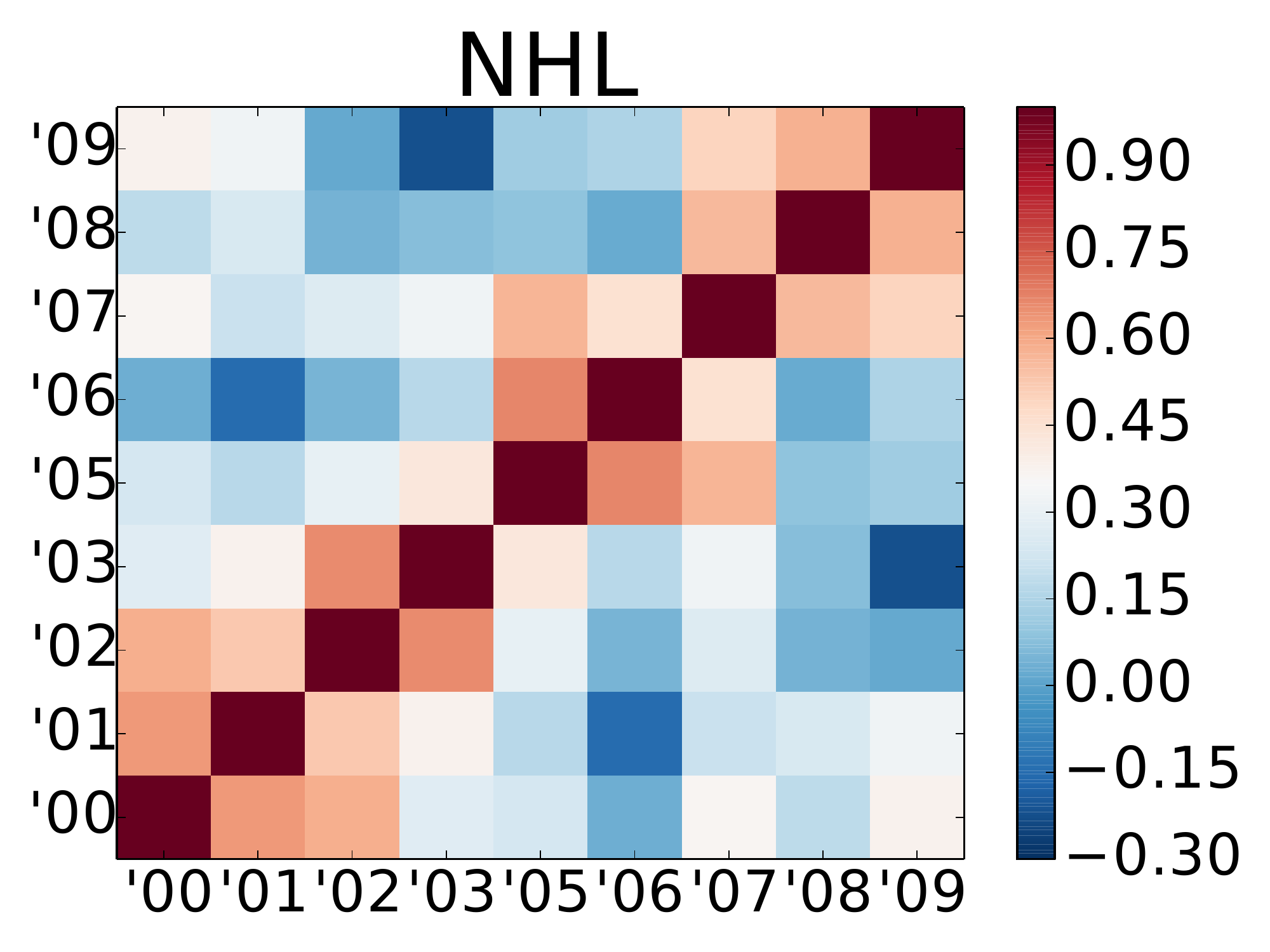}
  \caption{Correlation of inferred skills over years for each sport.  We see that the highest correlations occur along the block diagonal indicating that adjacent years are more similar.  Note that the scale is different for CFB due to a much higher correlation across all years.}
  \label{fig:skill_corr}
\end{figure}

Using the best overall model for each sport, we learn its parameters using all data in each particular season and calculate the Spearman rank correlation across team skills for each pair of seasons (Fig.~\ref{fig:skill_corr}). We find that the relative ordering of teams by their inferred skills exhibits strong serial correlation over time, which appears as a strong diagonal component in the pairwise correlation matrices. The low or inverse correlation in the far off-diagonal elements, as well as the block-like patterns observed in CFB and NFL, implies an underlying non-stationarity in team skills for each of the leagues over the roughly 10-year span of data.

The manner in which team rosters change over time is a likely source of such long-term dynamics in relative team skill. At short time scales, team rosters are fairly stable, with only a few players changing from season to season. However, over longer time scales, these changes accumulate, and rosters separated in time by more than a few years are likely to be very different, with concomitant differences in team skill.

The exception to this pattern is CFB, which shows a larger long-term correlation, i.e., a slower rate of change in relative team skills, than in professional sports. We speculate that this difference is caused by the difference in player mobility between college and professional-level sports:\ professional teams operate in a national player market, and players can move relatively freely among teams, while colleges operate as rough regional monopolies over the sources of their players.

\begin{figure*}
  \centering
  \includegraphics[width=\textwidth]{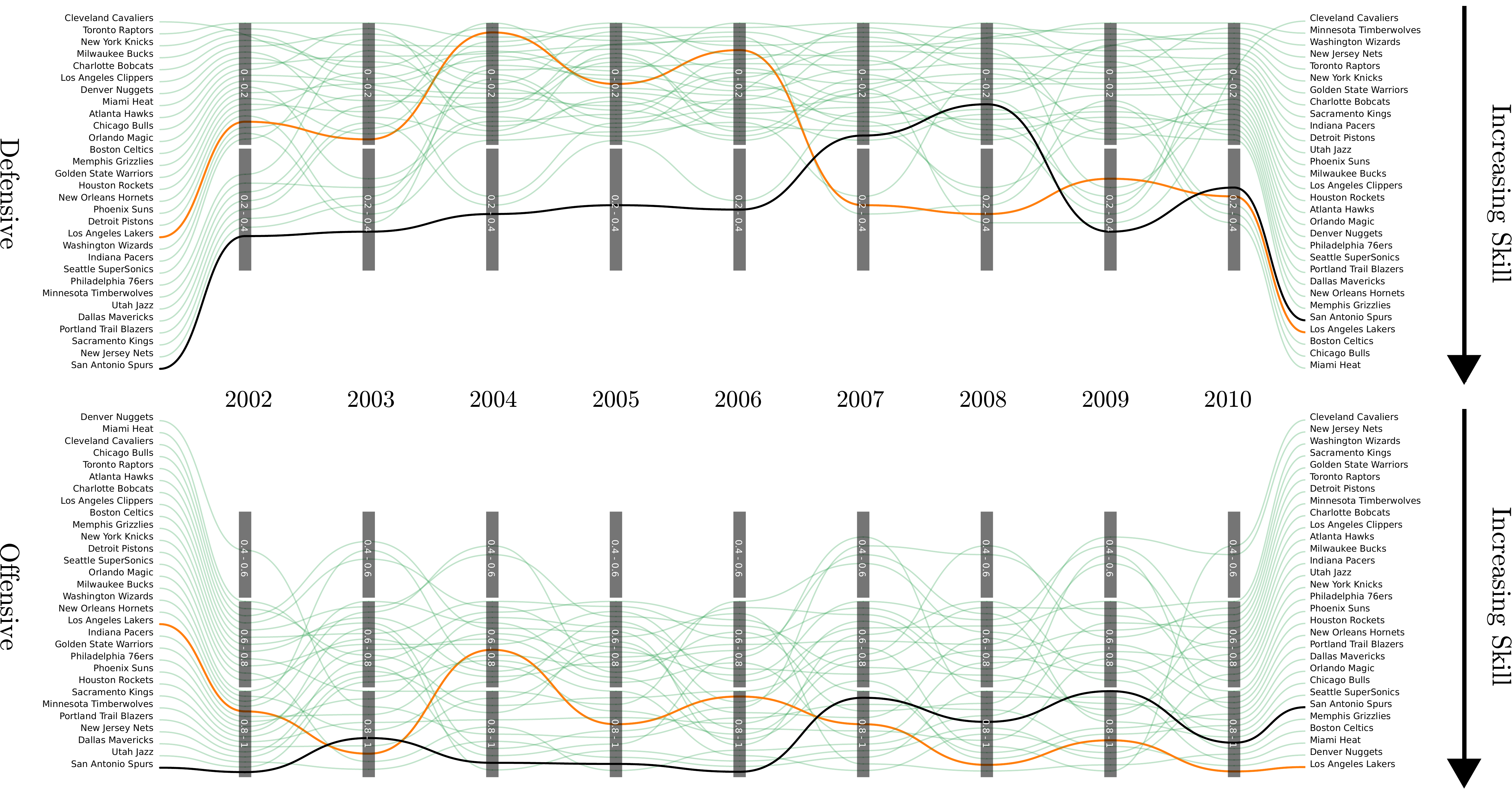}
  \caption{NBA defensive (\textit{top}) and offensive (\textit{bottom}) skill rankings.  Teams that won more than one NBA finals game in the data are highlighted, i.e., Lakers (\textit{orange}) 2002, 2009 and 2010, Spurs (\textit{black}) 2003, 2005 and 2007.}
  \label{fig:skillsnba}
  \vspace{3mm}
  \includegraphics[width=\textwidth]{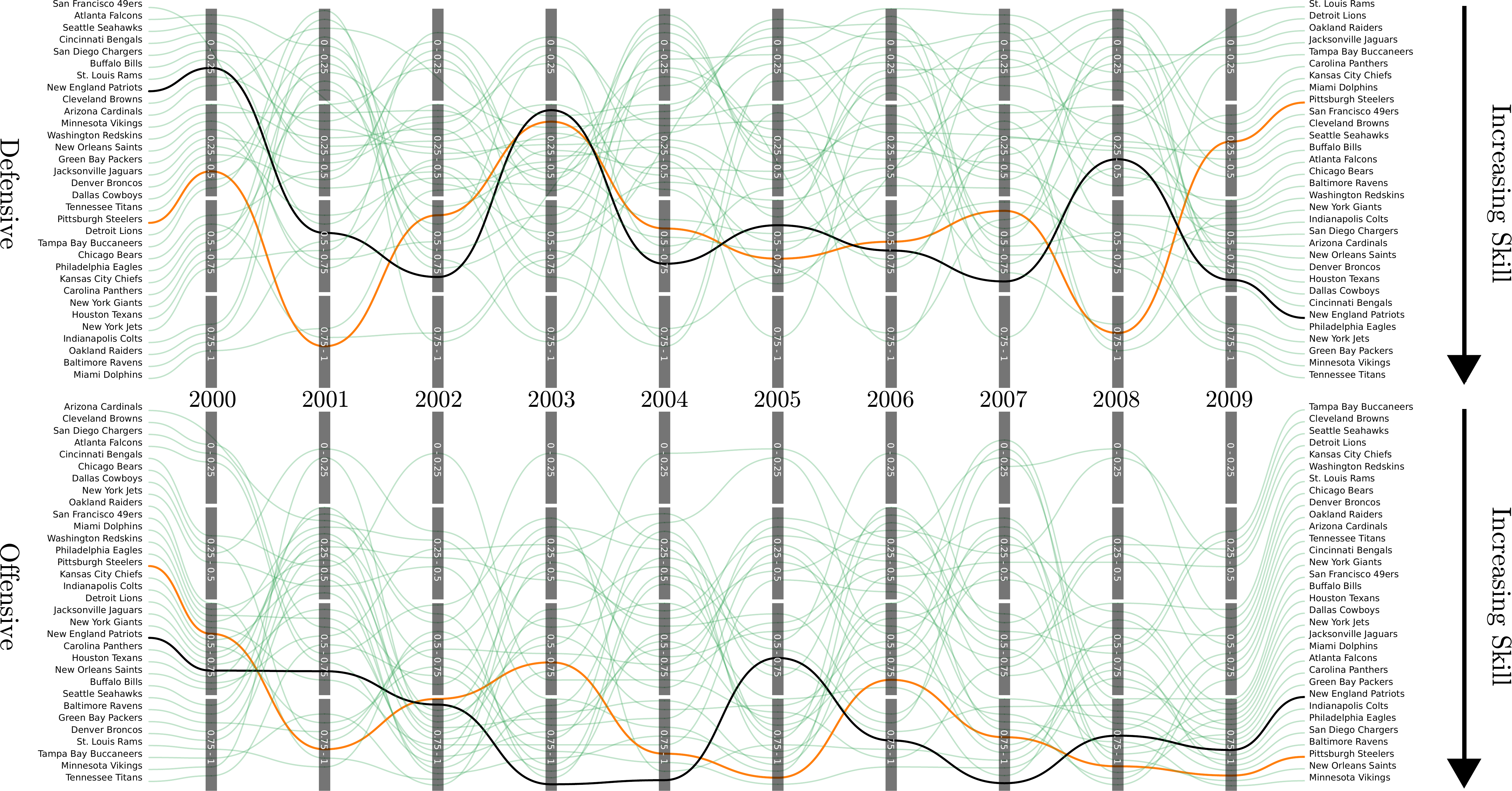}
  \caption{NFL defensive (\textit{top}) and offensive (\textit{bottom}) skill rankings.  Teams that won more than one NFL super bowl games in the data are highlighted, i.e., Patriots (\textit{black}) 2002, 2004 and 2005, Steelers (\textit{orange}) 2006 and 2009.}
  \label{fig:skillsnfl}
\end{figure*}

The inferred season-by-season skill orderings themselves are also of interest, as they reveal the particular trajectories of individual teams over time. We show visualizations of these trajectories for NBA and NFL in Figures~\ref{fig:skillsnba} and \ref{fig:skillsnfl}. We omit CFB because there are too many teams (461) to meaningfully visualize and NHL for space reasons.

For each plot we highlight the two teams that won the league championship (NFL Super Bowl and NBA Finals) more than once during the period covered by the dataset. It is notable that these teams are not necessarily the most skilled teams under our model. This is unsurprising, as tournaments by bracket are the highest variance method of identifying the most skilled team~\cite{ben2006choose}. Interestingly, in both NFL and NBA games, the highlighted teams tend to have strong offensive skills, while their defensive skills are more variable. This pattern suggests that offensive skills are more important for winning games, which seems reasonable given that a strong defense alone cannot win a game.

Looking at individual teams, we can see how their skills change with respect to the total ordering.  For instance, the Cleveland Cavaliers drafted LeBron James in 2003 and went from being ranked the third worst (offensive) team to a mid-range one. When James left for Miami Heat at the start of the 2010--11 season, we see the Cavaliers' offensive skill drop to the bottom ranked team, while Miami Heat's offensive and defensive skills increased to be ranked third and first respectively.   We also see that the Los Angeles Lakers' skills (both offensive and defensive) drop for the 2004--05 season.  After facing a difficult 2003-04 season~\cite{jackson2004last}, they disbanded the team, lost their coach and faced a number of injuries, resulting in a poorer performance in 2004--05.

Finally, the range of values occupied by offensive and defensive skills is different between NFL and NBA teams:\ in the latter, these two skills occupy non-overlapping ranges (large and small respectively), while in the former, they fall in similar ranges. That is, NBA teams are less likely to score when playing defensively than when they have possession of the ball, which serves to create a stronger anti-persistence scoring pattern (0.36) than for NFL (0.44), where skills are more evenly matched.

\section{Conclusion}
In this work we considered the online prediction tasks of \textit{Who will score next?} and \textit{Who will win?} based on the sequence of scoring events in the game so far.  Our probabilistic models based on latent team skills perform well at both predictive tasks and can predict with a high degree of certainty ($>80\%$) who will win a game in each of the four leagues we studied, after only half of the game has elapsed.  Furthermore, by using gameplay, i.e., the particular sequence of events within each game, to model team skill rather than just game outcomes, we can infer different types of latent team skills e.g., offensive vs.\ defensive skills.

Our statistical models provide a quantitative and principled means of capturing and testing hypotheses about the variability induced by chance, the biases produced by real differences in team skill, and the structural impact of game-specific rules. In applying these models to comprehensive data from four different sports, we found that each of the leagues is best fit by a different model.  This indicates that skill, luck, strategy, and the rules of the game serve different roles in the scoring dynamics across sports. The exception was professional hockey (NHL), where the very low scoring rate resulted in no clear predictive winner among our models.

Our models and results open up many new directions for future work. For instance, we could incorporate other data such as player or ball positioning~\cite{bourbousson2012space, yue2014learning}, or the timing of events~\cite{merritt2014scoring} to improve our models and allow us to apply them to low scoring sports such as soccer.  These models could also be used to make other predictions, e.g., the number of scoring events in a game, the final score, and when a lead is safe~\cite{clauset2015safe}, and to produce more rigorous team rankings (Figs.~\ref{fig:skillsnba} and~\ref{fig:skillsnfl}).

In addition to data on gameplay, data on individual player attributes and performance in competitive settings are also often available, e.g., height, strength, speed, accuracy when scoring, defensive skill, passing skill, etc. However, there are no good models that connect these characteristics to team skills and to gameplay as a means of predicting game outcomes.  The models we formulated here solve part of this problem by connecting team skill to gameplay. An interesting direction for future work would be to predict outcomes from player statistics via team skills as an intermediary.  Such a model would allow teams to make more data-driven choices about how they build team rosters and train their players. This extension of our work would also open up new avenues in designing realistic simulations of competitive play, e.g., for better AI in video games.

\section{Acknowledgements}
We thank Ruben Coen Cagli, Ramsey Faragher, Theofanis Karaletsos, Marina Kogan, David Edward Lloyd-Jones and Sam Way for helpful conversations, and acknowledge support from Grant \#FA9550-12-1-0432 from the U.S.\ Air Force Office of Scientific Research (AFOSR) and the Defense Advanced Research Projects Agency (DARPA).

\balance

\end{document}